\documentstyle[twoside,fleqn,npb,epsfig]{article}
\def\beq{\begin{equation}}
\def\eeq{\end{equation}}

\def\beqn{\begin{eqnarray}}
\def\eeqn{\end{eqnarray}}

\def\np#1#2#3  {{Nucl. Phys.~{\bf B#1} (19#3) #2}}
\def\nc#1#2#3  {{Nuovo. Cim.~{\bf #1} (19#3) #2}} 
\def\pl#1#2#3  {{Phys. Lett.~{\bf B#1} (19#3) #2}} 
\def\pr#1#2#3  {{Phys. Rev.~{\bf D#1} (19#3) #2}} 
\def\prl#1#2#3  {{Phys. Rev. Lett.~{\bf #1} (19#3) #2}} 
\def\prep#1#2#3 {{Phys. Rep.~{\bf #1} (19#3) #2}} 
\def\zp#1#2#3  {{Z. Phys.~{\bf C#1} (19#3) #2}} 
\def\rmp#1#2#3  {{Rev. Mod. Phys.~{\bf #1} (19#3) #2}} 
\def\JETP#1#2#3 {{Sov.\ Phys.\ JETP~{\bf #1} (19#3) #2}}
\def\sj#1#2#3 {{Sov.\ J.\ Nucl.\ Phys.~{\bf #1} (19#3) #2}}
\def\hepph  #1 {{\tt hep-ph/#1}}
\newcommand{\bfk}{\mbox{\boldmath $k$}}
\newcommand{\pup}{p^\uparrow}

\newcommand{\qup}{q^\uparrow}
\newcommand{\pdown}{p^\downarrow}

\newcommand{\qdown}{q^\downarrow}

\newcommand{\AmS}{{\protect\the\textfont2
  A\kern-.1667em\lower.5ex\hbox{M}\kern-.125emS}}
\title{
\vspace*{-1.5cm}
\begin{flushright}
ITP-SB-99-26 \\
June 1999 \\
\end{flushright}
Polarized lepton nucleon scattering --- \\ 
summary of the theory talks on spin physics at DIS 99}

\author{Werner Vogelsang
\address{Institute for Theoretical Physics, State University of New York
at Stony Brook, NY 11794-3840, U.S.A.}
}

\begin{document}

\begin{abstract}
We summarize the theory talks given in Working Group 4 `Polarized Lepton 
Nucleon Scattering' at the DIS 99 workshop. The significant progress
made over the last year on many of the interesting topics in `spin 
physics' is documented.
\end{abstract}

\maketitle
\section{INTRODUCTION}

Spin physics has been going through a period of great popularity
and rapid developments ever since the measurement of the proton's 
spin structure function $g_1$ by the EMC~\cite{emc} more than a decade ago. 
As a result of combined experimental and theoretical efforts, we have 
gained some fairly precise information concerning, for example, 
the quark spin contribution to the nucleon's spin. Yet, many other
interesting and important questions, most of which came up in 
the wake of the EMC measurement, remain unanswered so far, the most
prominent `unknown' being the fraction of the nucleon's spin 
carried by gluons. Future dedicated spin experiments (for a
compilation, see~\cite{dueren}) are expected to further improve our knowledge 
about this, and other, topics. 

We will organize this paper by first taking stock and summarizing some
main aspects of what has been learned about the nucleon's spin structure 
from experimental data on deep inelastic scattering (DIS) of polarized leptons 
and nucleons. Here we will mainly refer to 
talks~\cite{marco,stamenov,windmolders} given at this conference, in which 
next-to-leading order (NLO) QCD fits to the available DIS data were presented.

The second part of this paper will address the yet open questions in spin 
physics, most of which were discussed at this workshop. 
Some of the main issues here are the need for further information on the
nucleon's spin-dependent parton densities, in particular on its
gluon, the so far unmeasured transversity
densities, the orbital angular momentum carried by partons in the nucleon,
spin-transfer in fragmentation processes, and single-spin asymmetries.
Note that there is substantial overlap among some of these topics, so that a 
distinction between them in this work may at times look somewhat
artificial. On all these topics, and on others as well, significant and 
very recent theoretical advances were reported at this conference. 
In many cases, the progress made is of direct importance for future 
experimental studies.
\section{INFORMATION FROM POLARIZED DEEP INELASTIC SCATTERING}
Our present knowledge about the spin structure of the nucleon derives 
almost entirely from inclusive DIS of longitudinally polarized
leptons and nucleons. Here, experiments measure, for various targets, the 
spin asymmetry 
\begin{equation} \label{eq1}
A_1 (x,Q^2) \approx 
\frac{g_1 (x,Q^2)}{F_2 (x,Q^2)/[2 x(1+R(x,Q^2))]}  \; ,
\end{equation}
where $F_2$ is the unpolarized structure function and $R=F_L/2xF_1$. 
$g_1$ is directly related to the spin-dependent (anti)quark and gluon 
densities $\Delta q =q^+ - q^-$, $\Delta g =g^+ - g^-$. To NLO of QCD,
which currently is the `state-of-the-art' in the theoretical 
description, one has
\begin{eqnarray}
\label{eq2}
g_1 \!\!&=& \!\!  \frac{1}{2} \sum_q e_q^2 \left[
\left(\Delta q + \Delta \bar{q}\right) \otimes
\left(1+\frac{\alpha_s}{2\pi} \Delta C_q\right) \right. \nonumber \\
&&\!\!+ \left. \frac{\alpha_s}{2\pi} \Delta g \otimes \Delta C_g \right]\,\,,
\end{eqnarray}
where $\Delta C_q$, $\Delta C_g$ are the NLO terms in the coefficient 
functions. The polarized parton densities evolve in $Q^2$ according to the 
spin-dependent NLO DGLAP~\cite{dglap} equations. Note that beyond lowest 
order (LO),
the parton densities are not unique but refer to the factorization
scheme adopted in removing the collinear singularities occurring in the 
calculation of the NLO QCD corrections. This scheme dependence of
the parton densities is compensated by that of the coefficient
functions in such a way that a physical quantity like $g_1$ 
remains, to the order considered, 
unaffected by any change in the factorization scheme.
In other words, the physics content of phenomenological fits
to data has to come out the same, irrespective of the scheme 
chosen, if the analysis is performed properly.

Several NLO fits to the polarized DIS data were published
in the last few years~\cite{grsv,dss,smc}.
At this workshop, two updates of previous fits were 
presented~\cite{marco,stamenov}, using the most recently available
experimental information. It should be noted that inclusive DIS will in
principle only determine the sum of the polarized quark and antiquark 
distributions for each flavour, but not allow an immediate decomposition
into, say, sea and valence parts. Semi-inclusive measurements in 
DIS (see~\cite{dueren,windmolders}), which in principle could help out of this
situation, are presently not offering sufficiently precise information and
have (not yet) been routinely included in the global NLO fits
(see, however,~\cite{dss}). On the other hand,  the success 
of the parton model is based on the universality of parton densities,
and one certainly wants to use the information extracted from DIS 
to make predictions for other process, like polarized $pp$ reactions.
For such purposes, a full flavour decomposition is in general required.
Therefore, in order to end up with a complete set of parton densities, 
both~\cite{marco,stamenov} make further assumptions concerning the polarized 
quark sea. Even though the two fits differ slightly in some
aspects regarding, e.g., the input scale and densities for the
evolution, the main conclusions are the same and can be summarized as
follows:
\vspace*{-2mm}
\begin{itemize}
\item the fits work well, i.e., a very good quantitative
description of all data sets is achieved. This is demonstrated in~\cite{marco} 
Fig.~1;
\item the nucleon's singlet axial charge $a_0$ (directly related to the 
first moment of the flavour-singlet component of $g_1$) comes out to be 
only about $0.2$, implying (to LO) that quarks and antiquarks contribute 
only little to the nucleon spin. Thus, this
finding in the EMC experiment~\cite{emc}, which triggered so much 
theoretical and experimental activity over the last decade, 
has been confirmed over the years by all successive experiments (see 
also~\cite{windmolders});
\vspace*{-3mm}
\item the spin-dependent gluon density $\Delta g$ 
turns out to be only rather weakly constrained by the data. This 
is not a very surprising finding since, according to Eq.~(\ref{eq2}), 
$\Delta g$ enters $g_1$ only as a NLO correction, and of course indirectly 
via the $Q^2$-evolution of the parton densities. 
\end{itemize}

Note that on the last point the two fits~\cite{marco,stamenov} agree to a 
lesser extent; the error assigned to $\Delta g$ in~\cite{stamenov} is 
considerably smaller than in~\cite{marco} and probably does not fully 
reflect all existing uncertainties. The spread of allowed $\Delta g$ 
reported in~\cite{marco} is in line with that found in~\cite{smc}, where 
a very careful analysis of all uncertainties, experimental {\em and} 
theoretical, was performed.
\begin{figure}[t]
\centerline{
\vspace*{-1.1cm}
\epsfig{file=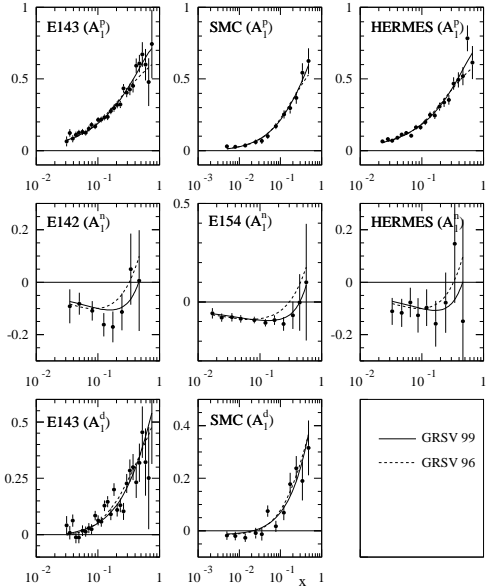,width=6.5cm} }
\vspace*{0.1cm}
\caption{Comparison of data on $A_1(x,Q^2)$ with a NLO
QCD fit~[3].}
\label{fig:fig1}
\vspace*{-0.4cm}
\end{figure}

An interesting possible explanation for the experimental smallness 
(with respect to constituent quark expectations) of $a_0$ was proposed 
in~\cite{bass} in terms of a non-perturbative gluonic `topology'
term arising from the non-invariance of the forward matrix elements 
of the Chern-Simons current $K^{\mu}$ under `large' gauge transformations
which change the topological winding number. Recall
that $a_0$ is given by the forward matrix element of the quark singlet
axial vector current $j^{\mu}_5$, and that $\partial_{\mu} j^{\mu}_5=
2 f\partial_{\mu} K^{\mu}$ is non-zero due to the axial anomaly.
This led~\cite{ar} to the interpretation of $a_0$ in terms
of the (conserved) quark spin and a gluonic contribution. The topology
term proposed in~\cite{bass} goes beyond this picture in the sense
that it induces an extra contribution to $a_0$ that has support only
at $x=0$. Consequently, it would have been missed in experimental
determinations of the first moment of $g_1$, which might explain the
finding of a small $a_0$. However, the topology term would contribute 
in determinations of $a_0$ in $\nu p$ elastic scattering, which implies that 
it can in principle be measured by comparing with the value for $a_0$ 
found in polarized DIS.

Two presentations at this workshop dealt with higher-twist contributions
to polarized structure functions. This is a very relevant issue 
since so far all experiments in polarized DIS have been fixed-target
experiments, with $Q^2$ values often not significantly higher than
a few GeV$^2$. In~\cite{avto}, target-mass corrections were calculated
for all polarized structure functions of neutral or charged current DIS.
In particular, the effects of such corrections on integral relations 
between polarized structure functions was systematically 
investigated, and new relations between the twist-3 parts of 
the spin-dependent structure functions were derived. First
numerical results indicate that the target-mass corrections
to $g_1$ are clearly non-negligible for $Q^2\approx 1$ GeV$^2$ and 
moderate $x$. The knowledge of such kinematical higher-twist terms
is also crucial if one wants to experimentally extract {\em 
dynamical} higher twists. Dynamical twist-4 contributions to the singlet part
of $g_1(x,Q^2)$ were estimated in~\cite{stein} in the 
context of the infrared renormalon model. Also from this model, 
non-negligible effects at experimentally relevant $x$ and $Q^2$ are
expected.
\section{OPEN QUESTIONS}
\subsection{Chasing $\Delta g$}
As explained above, the nucleon's spin-dependent gluon density $\Delta g$ 
is presently hardly constrained experimentally. Its presumably important 
r\^{o}le~\cite{ar} in understanding the nucleon spin has placed it in the focus
of attention of the spin physics community for a long time. Experiments
are being planned now (see~\cite{dueren}) which are dedicated to a 
large extent to a measurement of $\Delta g$. On the theory side, a lot
has been done, in particular in the past year, to provide the framework for
a reliable (perturbative) description of the key processes to be studied
in these experiments. The big improvement here has been the calculation
of NLO QCD corrections to the various reactions of interest. Clearly, only 
if the QCD corrections are under control, can a process that shows good 
sensitivity to $\Delta g$ at the lowest order level, be regarded as a 
genuine probe of the polarized gluon distribution and be reliably used to 
extract it from future data. Furthermore, it is
well-known that LO calculations of cross sections in $lN$
or $pp$ collisions are at best semi-quantitative. In particular, 
they are usually plagued by rather large uncertainties related
to the dependence of the theory prediction on the unphysical scales 
to be introduced in the process of renormalization of the strong
coupling and of factorization of mass singularities. In many
cases this situation is alleviated to a large extent when going to
the next order of perturbation theory. An example of this will be
shown below. 

To measure $\Delta g$ directly, it is crucial to consider processes
in which the gluon enters already at the lowest order, unlike in
the case of $g_1$. At the same time, one wants to pick processes
that have proven successful in the unpolarized case to constrain
the nucleon's unpolarized gluon density $g(x,Q^2)$.
Candidate processes, which will be studied experimentally in the 
future, are:

\vspace*{3mm}
\noindent
{\bf (a) Heavy-flavour production.}

\noindent
In leptoproduction, this process is to LO driven by  
photon-gluon fusion, $\vec{\gamma} \vec{g} \rightarrow h\bar{h}$, 
where, in practice, $h=c$ (the arrows indicate from now on a polarized 
particle). It presumably offers the cleanest signature for $\Delta g$ 
as far as fixed-target 
leptoproduction experiments are concerned. It has been one of the 
original motivations for the COMPASS experiment at CERN~\cite{dueren}, 
where one will actually look at charm {\em photo}production, i.e.
at the limit in which the exchanged photon is practically on-shell.
The calculation of QCD corrections to polarized heavy-flavour
photoproduction has been completed recently and was reported at this 
conference~\cite{bojak}. Note that not only the corrections to 
the Born channel are involved here, but also a new subprocess,
$\vec{\gamma} \vec{q}\rightarrow qc \bar{c}$, which is not present at LO. 
It turns out that, as expected, the QCD corrections are in general 
sizeable, in particular in the threshold and high-energy regions. 
However, they become reasonably small in the kinematical regime 
relevant for the COMPASS experiment. Extension to the case of
heavy-flavour production by two polarized protons, is nearing completion.
Here, the main LO contribution results from the partonic channel
$\vec{g}\vec{g}\rightarrow c\bar{c}$, which obviously has strong 
sensitivity to $\Delta g$. The NLO corrections to this process still
need to be completed, while those for the other channel,
$\vec{q}\vec{\bar{q}}\rightarrow c\bar{c}$, as well as the new NLO
process $\vec{q}\vec{g}\rightarrow c\bar{c}q$, are already known~\cite{bojak}.
The full QCD corrections to $\vec{p}\vec{p}
\rightarrow c\bar{c}X$ will be of great relevance for heavy-flavour 
experiments to be carried out at the polarized RHIC collider at 
BNL~\cite{dueren}.

The NLO QCD corrections to $\vec{\gamma}\vec{g} \rightarrow c\bar{c}$ with a 
{\em virtual} photon, i.e. the corrections to the charm component $g_1^c$ 
of the structure function $g_1$, are almost complete as well~\cite{jack}.
First numerical results indicate that the relative contribution of 
$g_1^c$ to the total $g_1$ is smaller than that of $F_2^c$ to $F_2$ 
in the unpolarized case. This is readily understood from the fact
that there are oscillations of the relevant coefficient functions
due to the sum rule $\int dz H_g^S(z)=0$ for charm production off an 
initial polarized gluon~\cite{jack}.

\vspace*{3mm}
\noindent
{\bf (b) Polarized photoproduction of jets.}

\noindent
The process $\vec{e} (\vec{\gamma}) \vec{p} \rightarrow jet(s) X$ 
would be accessible experimentally~\cite{butter} after an upgrade of the HERA
collider to having a polarized proton beam in addition to 
the already polarized electron beam. Again, it has a LO contribution
sensitive to $\Delta g$: $\vec{\gamma} \vec{g} \rightarrow q\bar{q}$.
Complications arise due to the fact that, as is well-known from the
unpolarized case, the photon will not only interact in a `point-like'
(direct) way, but also via its partonic structure (`resolved' component). 
Nothing is known so far experimentally about the spin-dependent parton 
densities $\Delta f^{\gamma}$ of the polarized photon. On the one hand, this
limits our predictive power for this process; on the other hand, it makes 
the process and its experimental study even more fascinating, since 
a measurement of the $\Delta f^{\gamma}$ in future photoproduction experiments
at polarized HERA appears feasible~\cite{butter}.

NLO QCD corrections are expected to be particularly 
important for the case of jet production, since it is only at NLO that the 
QCD structure of the jet starts to play a r\^{o}le in the theoretical 
description, providing for the first time the possibility, and necessity,
to realistically match the experimental conditions imposed to define a jet.
In~\cite{daniel} the calculation of the QCD corrections to polarized 
single-inclusive photoproduction of jets, both for the `direct' and the 
`resolved' components, was reported. The corrections are found to be 
particularly important in the region of negative rapidities of the jet, 
where the `direct' component dominates. Also, the expected strong 
reduction in scale dependence of the cross section when going from LO to NLO 
was verified. 

\vspace*{3mm}
\noindent
{\bf (c) Jet production in polarized $pp$ collisions.}

\noindent
This process presumably offers one of the most promising ways to
measure $\Delta g$ at RHIC. It combines a potentially strong sensitivity to 
$\Delta g$, thanks to the partonic LO channels $\vec{q} \vec{g} \rightarrow 
qg$ and $\vec{g}\vec{g}\rightarrow q\bar{q}(gg)$, with a large event rate,
which results in rather small expected statistical errors on the corresponding
spin asymmetry. Indeed, this is exactly what is found in the theory
studies presented~\cite{daniel} here. Fig.~2 shows the asymmetry 
for single-inclusive jet production for various sets of 
spin-dependent parton densities, which differ mainly in the size
of $\Delta g$. In addition, the smallest jet asymmetry that can be measured
experimentally at RHIC for ${\cal L}=100$ pb$^{-1}$ is indicated by the 
open blocks. It is obvious that jet measurements at RHIC will be in the 
position to pin down $\Delta g$. Fig.~3 demonstrates the significant 
reduction in scale dependence at NLO, which makes the theory predictions
much more reliable.
\vspace*{-0.6cm}
\begin{figure}[ht]
\hspace{-.3cm}
\centerline{
\epsfig{figure=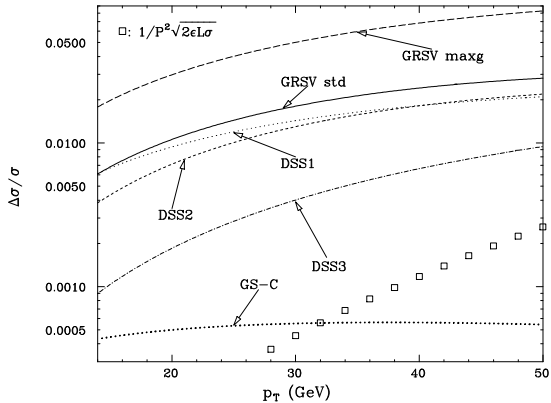,width=0.45\textwidth,clip=} }
\vspace{-0.8cm}
\caption{NLO single-inclusive jet asymmetries in $\vec{p}\vec{p}$ 
collisions at RHIC, for various polarized parton densities~[17].}
\vspace{-0.4cm}
\end{figure}                                                              
\begin{figure}[h]
\vspace{-1.2cm}
\centerline{
   \epsfig{figure=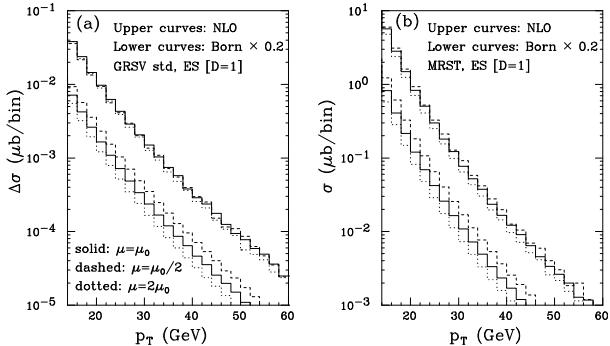,width=0.5\textwidth,clip=} }
\vspace{-0.8cm}
\caption{Scale dependence of the NLO and LO jet $p_{T}$-distributions in 
$\vec{p}\vec{p}$ collisions~[17].  }
\vspace{-0.4cm}
\end{figure}                                                              

\vspace*{-3mm}
\noindent
{\bf (d) Prompt photon production.}

\noindent
The production of high-$p_T$ `prompt' photons in unpolarized hadron-hadron 
collisions has historically been considered as {\em the} probe of the hadrons'
gluon density. Data have been the backbone of the gluon determination
in many analyses of parton densities. This is thanks to the presence 
and dominance of the LO `Compton' process $qg\rightarrow \gamma q$ and
the alleged `cleanliness' of the photon final state. These features 
make prompt photon production in $\vec{p}\vec{p}$ collisions at RHIC the 
most promising candidate for a measurement of $\Delta g$. 
However, there are caveats: firstly, the process is not as clean as it 
might look superficially:
there is also a contribution for which a final-state quark or gluon,
produced in a hard pure-QCD subprocess, {\em fragments} into a photon.
The photon fragmentation functions are only insufficiently
known at present, which sets a limitation to our predictive power 
for the prompt photon process. The situation is alleviated somewhat 
by the fact that in prompt photon experiments at colliders, an `isolation 
cut' is imposed on the photon in order to reduce the huge background 
from the copiously produced $\pi^0$. Traditionally, this is defined by 
drawing a cone of fixed aperture around the photon, and by restricting the 
hadronic energy allowed in this cone. In this way the fragmentation 
component, which results from an essentially collinear process, is 
diminished; however it still remains non-negligible.
A way out of this dilemma was proposed at this conference~\cite{stefano}: 
by refining the isolation cut to the extent that one allows
{\em less and less} hadronic energy the closer to the photon it is
deposited, until eventually {\em no} energy at all is allowed exactly 
collinear to the photon, it is 
possible to define an isolated prompt photon cross section that is 
entirely independent of any fragmentation component. This definition
of isolation has therefore a clear advantage over the traditional one.
It should be possible to implement it in experiment without too much effort.
Phenomenological applications at RHIC energies look very 
promising~\cite{stefano} as far as the possibilities of measuring
$\Delta g$ are concerned.

A potentially more hazardous problem concerning the prompt photon 
process at RHIC is that
in the unpolarized case the agreement between theory and some of the 
most precise data sets~\cite{cdf} is rather poor and sometimes 
even appallingly bad. Clearly, if this situation persists, one 
will have to worry whether in the polarized case one can really 
interpret future data straightforwardly in terms of $\Delta g$. Without going 
any further into the discussion on the situation in the unpolarized case, 
it should be emphasized that in view of the present experimental and,
in particular, theoretical efforts~\cite{cmnov} in this field, it seems
likely that the theoretical description of prompt-photon production
will be in much better shape by the time RHIC will deliver first data 
on $\vec{p}\vec{p}\rightarrow \gamma X$.
\subsection{Transversity and Drell-Yan dimuon production} 
Apart from the unpolarized and longitudinally polarized parton densities,
a third type of twist-2 parton distribution of the nucleon can be 
defined, the `transversity' density~\cite{rs,am} $\Delta_T q(x,Q^2)$. In a 
transversely polarized nucleon it counts the number of quarks with spin 
aligned parallel to the nucleon spin minus the number of quarks with opposite 
polarization. Unfortunately, nothing is known as yet experimentally about 
the transversity distributions. Technically speaking, they are `chiral-odd', 
that is, a helicity-flip of the quark-line is required in the hard 
process~\cite{am} that probes the $\Delta_T q$. This property makes 
the transversity densities inaccessible in fully inclusive DIS. 
Drell-Yan lepton pair production in collisions of transversely polarized 
protons offers a possibility to get access to the $\Delta_T q$ 
\cite{rs,am}, and the RHIC spin physics program comprises experiments of 
this kind.

A `positivity' inequality for the $\Delta_T q (x,Q^2)$ has been derived 
in~\cite{soffer1} in terms of the unpolarized and longitudinally polarized
quark densities: $|\Delta_T q(x,Q^2)|\leq (q(x,Q^2)+\Delta q(x,Q^2))/2$,
which may be used for modeling the $\Delta_T q$ when making
predictions for future experiments. This inequality is preserved
by NLO DGLAP evolution in the ${\overline{\rm MS}}$ scheme, as discussed
in~\cite{teryaev} in terms of a `gain-loss' equation.

Interesting information on the tensor charge of the nucleon, which is
related to the first moment combination $\delta q\equiv \int_0^1 dx
(\Delta_T q-\Delta_T \bar{q})(x,Q^2)$,
was recently obtained in lattice calculations~\cite{horsley}. It turns 
out that the quantity $\delta u-\delta d$, after conversion
to a renormalization-group invariant, is rather similar to
the corresponding non-singlet contribution of $u$ and $d$ flavours to the 
proton's {\em axial} charge on the lattice; see~Fig.~4. Note though that 
the latter should come out as $g_A\approx 1.26$, which does not
happen satisfactorily, presumably as a result of the quenched approximation
used in the calculation~\cite{horsley}.
\begin{figure}[t]
\vspace{0.6cm}
\hspace*{4mm}
\epsfxsize=6.50cm \epsfbox{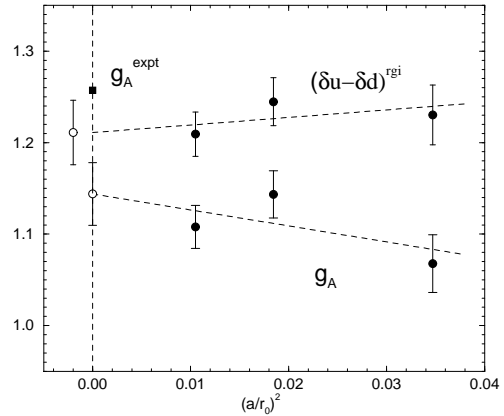}
\vspace{-0.9cm}
\caption{The continuum limit for $g_A$ and
$(\delta u - \delta d)^{rgi}$~[25].}
\vspace{-1cm}
\end{figure}   

Phenomenological NLO predictions for the Drell-Yan process with
transversely polarized protons were presented at this workshop 
in~\cite{oliver,miyama}. In~\cite{oliver}, {\em saturation} of Soffer's 
inequality at a low input scale for the evolution was assumed in order to
model the transversity densities. With some justification, this approach can
be expected to predict the maximally possible spin asymmetries
for the Drell-Yan process. These asymmetries turn out to be of the
order of a few per cent. The expected statistical errors for  
such measurements at RHIC or HERA-$\vec{\rm{N}}$ turn out to be 
smaller than the asymmetry by a factor of 2$-$3 in the 
optimal case; see Fig.~5.
A possible flavour ($\Delta_T \bar{u}$, $\Delta_T \bar{d}$) separation of the 
transversity sea in dimuon production was discussed in~\cite{miyama},
the proposed tool for this being the ratio $\Delta_T \sigma^{pd}/2 
\Delta_T \sigma^{pp}$ of Drell-Yan cross sections; however, the experimental 
feasibility appears unlikely, also in view of the findings of~\cite{oliver}.

The Drell-Yan process also was the main topic in~\cite{koike,kumano}.
In~\cite{koike}, the production of dimuon pairs was studied for the
case of `LT'-collisions of transversely polarized protons with longitudinally
polarized ones. The description involves interesting twist-3 structure 
functions. Employing bag model predictions and large-$N_C$ evolution for the 
latter, estimates for the corresponding 
spin-asymmetry $A_{LT}$ are obtained which are
much smaller than those for $A_{TT}$ found in~\cite{oliver}.
Many new, yet unexplored, structure functions appear if the Drell-Yan 
process is studied in polarized proton-deuteron scattering, thanks to the 
spin-1 nature of the deuteron~\cite{kumano}.
\begin{figure}[t]
\vspace{-0.3cm}
\centerline{
   \epsfig{figure=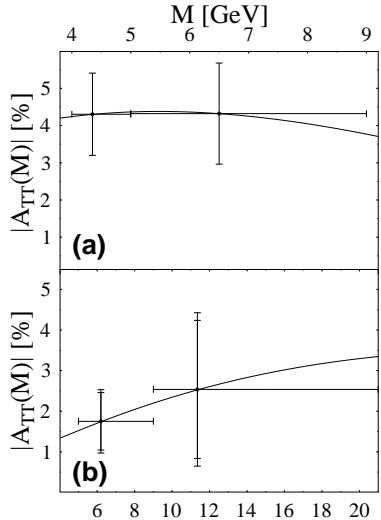,width=5.5cm,clip=} }
\caption{The `maximal' double-spin asymmetry for the Drell-Yan process in 
collisions of transversely polarized protons at (a) 
HERA-$\vec{\rm{N}}$ and (b) RHIC, as a function of the 
dimuon mass $M$~[26].} 
\vspace{-0.8cm}
\end{figure}   
\subsection{Orbital angular momentum, DVCS, and off-forward parton 
distributions}
It will only really be justified to refer to the spin structure of 
the nucleon as `understood' if also the contributions resulting from the 
orbital angular momenta $L_{q,g}$ of quarks and gluons will have been 
determined. As was discovered in~\cite{ji}, the {\em total} angular momenta 
(i.e., integrated over all Bj\o rken-$x$)
carried by quarks and gluons, $\Delta \Sigma/2 +L_q$ and $\Delta G +L_g$, 
respectively, can in principle be determined 
in `deeply-virtual Compton scattering' (DVCS) $\gamma^* P
\rightarrow \gamma P'$: the explicit presence
of $x^{\mu}$ in the corresponding terms of the QCD angular momentum tensor
forces one to go to the off-forward region~\cite{jaffe}.
Things become more complicated if one considers `parton distributions'
of orbital angular momentum, i.e. $L_{q,g}$ as functions of Bj\o rken-$x$.
As pointed out in~\cite{jaffe}, one encounters the dilemma here that
such quantities can indeed be consistently defined~\cite{jafrefs},
but that presently no way of observing them in experiment has been 
discovered. Vice versa, quantities that {\em can} in principle be measured 
in DVCS, do not directly find their interpretation in terms of 
$L_{q,g}(x)$~\cite{jaffe}. 

The DVCS amplitude itself was discussed in several talks at this workshop.
In~\cite{olness}, its asymptotic properties in the very forward 
and backward regions were studied, resumming the large logarithms arising.
The most general DVCS process, $\gamma^* P \rightarrow \gamma^{*'} P'$
was examined in~\cite{robaschik} in a `generalized' Bj\o rken 
region with $-(q_1+q_2)^2/4\nu$ and $(q_1^2-q_2^2)/2\nu$ as scaling
variables, $q_1$, $q_2$ being the photon momenta. The technical
tool employed here is the `nonlocal light-cone expansion', which
was also the motivation for~\cite{geyer}, where a way of systematically 
decomposing nonlocal light-cone operators into contributions of 
definite twist was presented. In the forward case the analysis
of~\cite{robaschik} recovers the known integral relations between polarized
DIS structure functions.

Being generalizations of the usual parton densities, off-forward parton 
distributions break Bj\o rken scaling as soon as QCD radiative corrections
are taken into account. The resulting evolution equations were studied 
in~\cite{belitsky}, where even the {\em next-to-leading} order 
anomalous dimensions were included. Remarkably, a way of deriving these
was found that avoids the involved calculation of explicit two-loop
diagrams, but makes clever use of conformal covariance, the known
two-loop anomalous dimensions for the forward case, and the recognition
that the breaking of conformal invariance (which takes place at the
NLO level) can be accounted for by calculating {\em one-loop}
conformally anomalous terms. First numerical results in the non-singlet
sector indicate that the NLO terms induce only very mild corrections to
LO evolution; however, as with many things, the calculation
has to be done before one can tell! In~\cite{mueller}, a complete set of 
${\cal N}=1$ supersymmetric constraints for the anomalous dimensions of 
conformal twist-2 operators was presented. As an application, the ER-BL 
evolution~\cite{erbl} kernel at NLO in the parity-odd sector, which
governs both the evolution of the corresponding singlet meson 
distribution amplitude and of the off-forward distribution, was derived 
without any evaluation of a two-loop diagram.
\subsection{Spin-dependent fragmentation functions}
Studies of spin-transfer reactions provide insight into the 
field of `spin physics' from a different angle. Rather than asking
for the `spin-up minus spin-down' distribution of, say, a 
quark in a polarized proton, one is primarily interested here in
the fragmentation counterpart, i.e. the distribution of a polarized 
(spin-$\frac{1}{2}$) hadron $h$ in a polarized parton $f$, 
\begin{equation}
\label{eq:deltad}
\Delta D_f^h(z,Q^2) \equiv D_{f(+)}^{h(+)}(z,Q^2) - D_{f(+)}^{h(-)}(z,Q^2)\; .
\end{equation}
Obviously, access to such quantities is only possible if one is able to 
measure the polarization of $h$. $\Lambda$-baryons are particularly suited for
such studies due to the self-analyzing properties of their dominant weak
decay $\Lambda \rightarrow p \pi^-$. Recent LEP measurements~\cite{data} 
of the polarization of $\Lambda$'s produced in $e^+e^-$ annihilation, 
have demonstrated the experimental feasibility of reconstructing 
the $\Lambda$ spin. Here, {\em{no}} polarization of the initial beams is 
required since the parity-violating $q\bar{q}Z$ coupling automatically 
generates a net polarization of the quarks that fragment into the 
$\Lambda's$.

At the moment, the sparse data available from LEP are by far not sufficient to 
constrain the $\Delta D_f^{\Lambda}$ significantly. Various reactions 
that might be employed for obtaining independent information have
been discussed in~\cite{soffer} at this workshop, and phenomenological
predictions for them were presented, based on conceivable sets
of $\Delta D_f^{\Lambda}$ proposed in~\cite{dsv}. These include 
semi-inclusive DIS, $\vec{e}p\rightarrow \vec{\Lambda}X$, and in
particular $\vec{p}p \rightarrow \vec{\Lambda}X$. The latter reaction
could be studied at RHIC, and the prospects are very promising as 
far as the expected statistical errors are concerned in comparison
to the actual possible size of the spin-transfer asymmetry. This holds true
for longitudinal as well as for transverse polarization of the proton beam 
and the $\Lambda$. Finally, interesting information on the flavour 
decomposition of the $\Delta D_f^{\Lambda}$ could also be obtained in 
$\nu$, $\bar{\nu}$ semi-inclusive DIS~\cite{soffer}.
\subsection{Transverse single-spin asymmetries}
The understanding of single-spin asymmetries for processes like
$p^{\uparrow} p\rightarrow \pi X$, $l p^{\uparrow} \rightarrow \pi X$, etc.
($\uparrow$,$\downarrow$ denoting transverse polaj9iysrization)
is another interesting challenge. Experimentally, the study of 
such asymmetries is simpler than that of double-spin asymmetries, 
for obvious reasons. On the theoretical side, within the `normal' framework 
of perturbative QCD and the factorization theorem at leading twist for 
collinear parton configurations, no non-zero single-spin
asymmetry can be constructed for parity-conserving processes. 
However, it has been recognized for a while now~\cite{sivers,collins,boer} 
that possible origins of single-spin asymmetries may be found in the 
dependences of parton distribution and fragmentation functions 
on intrinsic parton transverse momentum $k_T$. Under the crucial 
assumption that a factorization theorem can be invoked even when 
not integrating out $k_T$, one can rather successfully set up a 
hard-scattering model to describe the single-spin asymmetries in terms
of the usual convolutions of perturbatively calculable partonic 
hard-scattering cross sections with various, now $k_T$-dependent, parton 
densities and fragmentation functions~\cite{anselmino,boer}. The 
set of the latter is bigger than the one of the usual ($k_T$ integrated)
distributions considered in the previous sections. For instance, one may
have~\cite{anselmino,boer},
in fairly obvious notation,
\begin{eqnarray} \label{fcts}
\Delta^N D_{h/q} =
D_{h/\qup}(z, \bfk_{\perp}) -
D_{h/\qdown}(z, \bfk_{\perp})\;,  \nonumber \\
\Delta^N f_{q/\pup} =
f_{q/\pup}(x, \bfk_{\perp}) -
f_{q/\pdown}(x, \bfk_{\perp})\;,  \nonumber \\
\Delta^N f_{\qup/p} =
f_{\qup/p}(x, \bfk_{\perp}) -
f_{\qdown/p}(x, \bfk_{\perp})\; ,
\end{eqnarray}
all of which integrate to zero over $k_T$ and are $T$-odd.
For the single-spin asymmetry in $p^{\uparrow} p\rightarrow \pi X$,
all of the three functions in~(\ref{fcts}) might be at work and 
correspond to the Collins 
effect~\cite{collins}, Sivers effect~\cite{sivers}, and the effect
proposed in~\cite{boer}, respectively. There is a qualitative
difference between the first mechanism and the other two in that, 
in order to be able to produce an effect, the latter rely on the presence 
of some kind of initial-state interactions between the incoming particles,
whereas the Collins effect just requires {\em final}-state interactions (which 
are certainly present), to make the overall process time-reversal symmetry
conserving. This makes the Collins effect perhaps a more likely
source for single-spin asymmetries in general, and definitely for
reactions with a lepton-nucleon initial state. In consequence,
in~\cite{anselmino} the present data for the single-spin asymmetry
in $p^{\uparrow} p\rightarrow \pi X$ were fitted just in terms of 
the Collins effect, and the resulting $D_{h/\qup}(z, \bfk_{\perp})$
was then used to predict the asymmetry in semi-inclusive single-spin
DIS, $l p^{\uparrow} \rightarrow \pi X$. Large asymmetries
are found for HERMES kinematics; see~Fig.6. 
Note though that the $p_T$ of the pion
considered here is only just larger than 1 GeV, which raises some concern 
as to whether the hard-scattering mechanism employed is fully appropriate.
\begin{figure}[h]
\vspace{-0.6cm}
\centerline{
   \epsfig{figure=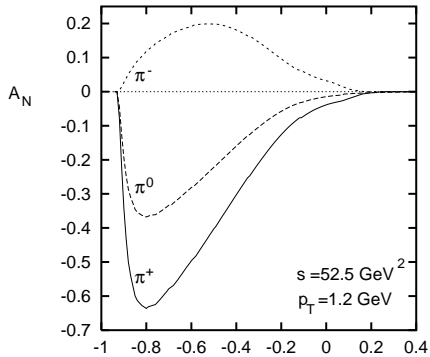,width=5cm,angle=270,clip=} }
\vspace{-0.9cm}
\caption{Predictions for the 
transverse single-spin asymmetry $A_N(lp^{\uparrow}
\rightarrow \pi X)$ as a function of Feynman-$x_F$ for typical 
HERMES kinematics~[45].}
\end{figure}   

Distributions similar to the first two in~(\ref{fcts}) 
were considered in~\cite{boglione}, where $k_T^2$-weighted integrals 
of the functions were used to study azimuthal asymmetries in DIS, like the one
presented by the HERMES collaboration at this workshop (see~\cite{dueren}).

The Collins effect has been studied using LEP data on $Z^0 \rightarrow 2 
{\rm jet}$ decay~\cite{efremov}. Here one exploits that the transverse 
polarizations of the produced quark and antiquark are correlated as 
$(v_q^2-a_q^2)/(v_q^2+a_q^2)$. The magnitude of the Collins effect 
is represented by the analyzing power~\cite{efremov,bjm}
$A\equiv|\Delta^N D_{\pi/q}|/D_{\pi/q}$, i.e. the ratio of the fragmentation
function introduced in~Eq.(\ref{fcts}) over the unpolarized one.
Preliminary results suggest $A$ to be clearly non-zero.

As was shown in~\cite{boer}, the function $\Delta^N f_{\qup/p}$ in~(\ref{fcts})
(named $h_1^{\perp}$ there) also offers a possible explanation for an 
anomalously large $\cos(2 \phi)$ dependence of the unpolarized Drell-Yan cross
section found by the NA10 collaboration~\cite{na10}. Note that
for the Drell-Yan process the Collins effect is clearly not involved. 
Furthermore, {\em single-spin} Drell-Yan measurements at RHIC are 
sensitive to the convolution $\Delta^N f_{\qup/p} \otimes \Delta_T q$ 
and might therefore be useful to constrain the transversity 
densities~\cite{boer}.
\subsection{Two further topics in spin physics}
The spin structure of the {\em photon} is another interesting topic
in spin physics, about which no experimental information is 
available yet. The case of polarized (quasi)real photons was
studied a few years ago~\cite{gsv}, and it was shown that the
corresponding polarized parton densities would be accessible
in jet-photoproduction experiments at a polarized HERA collider~\cite{butter}.
At this conference, predictions for the structure function
$g_1^{\gamma}(x,Q^2,P^2)$ of a photon that is off-shell by $-P^2$, were
presented~\cite{uematsu}. Only the perturbative (`point-like') part was 
considered. The first moment, $\int_0^1 dx g_1^{\gamma} 
(x,Q^2,P^2)$, comes out non-vanishing, unlike in the case of real photons.

The possibilities of analyzing the chiral structure of various
conceivable leptoquark models using a polarized HERA collider
or even a polarized TESLA$\otimes$HERA machine were examined 
in~\cite{virey}. The tool to do this would be the parity-violating
DIS asymmetry $A^{PV}=(\sigma_{NC}^{--}-\sigma_{NC}^{++})/
(\sigma_{NC}^{--}+\sigma_{NC}^{++})$, where superscripts denote
electron and proton helicities. For an unambiguous separation of the
various models, polarized $en$ collisions would be required in addition
to $ep$ ones.

\vspace*{0.2cm}
\noindent
{\bf Acknowledgments:}
It is a pleasure to thank all participants of our working group 
for their contributions and for many interesting discussions. I am also 
grateful to the organizers of `DIS 99', in particular to J.\ Bl\"{u}mlein,
for their tireless efforts and support. My thanks also go to M.\ D\"{u}ren
for his collaboration in organizing our working group.

\vspace*{0.2cm}
\noindent
Note concerning the references: For most talks we just refer to the 
contribution to the proceedings, and mention only the author who actually 
presented the talk. For the other authors and/or for previously 
published papers on the same topic by the author(s), please 
consult their actual 
contribution.
\normalsize

\end{document}